\documentclass[twocolumn,aps,showpacs,floatfix,superscriptaddress]{revtex4}
\usepackage{amsmath,amssymb,eucal,graphicx}

\def\av#1{\langle#1\rangle}
\def\g{{\bf g}}
\def\A{{\bf A}}
\def\tlam{{\tilde\lambda}}

\begin{document}
\title{A Toy Model of the Rat Race}
\author{D. ben-Avraham}
\email{qd00@clarkson.edu}
\affiliation{Department of Physics, Clarkson University, Potsdam, New York 13699 USA}
\author{Satya N. Majumdar}
\email{majumdar@lptms.u-psud.fr}
\affiliation{Laboratoire de Physique Th\'eorique et Mod\`eles Statistiques
  (UMR 8626 du CNRS), Universit\'e Paris-Sud, B\^atiment 100, 91405 Orsay
  Cedex, France}
\author{S.~Redner}
\email{redner@bu.edu}
\affiliation{Center for Polymer Studies and Department of Physics, Boston University, Boston,
Massachusetts 02215 USA}

\begin{abstract}
  We introduce a toy model of the ``rat race'' in which individuals try to
  better themselves relative to the rest of the population.  An individual is
  characterized by a real-valued fitness and each advances at a constant rate
  by an amount that depends on its standing in the population.  The leader
  advances to remain ahead of its nearest neighbor, while all others advance
  by an amount that is set by the distance to the leader.  A rich dynamics
  occurs as a function of the mean jump size of the trailing particles.  For
  small jumps, the leader maintains its position, while for large jumps,
  there are long periods of stasis that are punctuated by episodes of
  explosive advancement and many lead changes.  Intermediate to these two
  regimes, in a typical realization of the system, agents reach a common
  fitness and evolution grinds to a halt.

\end{abstract}
\maketitle

\section{INTRODUCTION}

A basic fact of life is competition.  In evolution, only the fittest survive;
in the workplace, we compete for professional advancement; in social events,
we compete for attention; in sports, its very purpose is to excel in
competition.  Idealized models of social competition have recently been
proposed in which the status of each individual is determined by competitive
success \cite{btd,br,bvr,msk}.  In this spirit, we introduce a simple
``rate-race'' model that embodies the struggle for advancement in a
competitive environment.  Because everyone is engaged in the same perpetual
rat race, one's relative standing may change slowly or not at all, even
though the population as a whole may be advancing.  When the competition
favors the strong, the leader runs away from the rest of the population.  As
the competition becomes more equitable, in any typical realization of the
system, everyone reaches the same fitness and the population become static.
When the leader is easily overtaken, the mean fitness undergoes periods of
near stasis and explosive advancement that qualitatively mirrors the
phenomenon of punctuated evolution \cite{punc}.

Empirical motivations for our model come from evolution and from sports.  In
evolution, large-scale species extinctions occur during sudden spurts, with
much slower development during the intervening periods \cite{punc,evol}.
These periods of near stasis characterize many sports, where it is not
possible to maintain a long-term competitive advantage.  If one finds such a
winning strategy, competitors will eventually find a counter-strategy so that
any advantage is lost.  Conversely, a consistent loser will be replaced by a
more competent individual so that losing strategies also do not persist.

A famous example of the latter idea comes from baseball, where the mythic
achievement of a .400 hitter, an exceptional player who gets a hit in more
than 40\% of his turns at bat, occurred multiple times during the early years
of the sport---25 times from 1871 to 1941 (last accomplished by the .406
batting average of Ted Williams of the Boston Red Sox in 1941)---but none
since then.  An appealing explanation for this phenomenon, proposed by S. J.
Gould \cite{G}, is that the increasing competitiveness as the sport has
developed makes outliers less likely to occur.  To illustrate this point,
Gould found that the dispersion in the batting averages of all regular
players decreased systematically from 1875 until 1980, even though
year-to-year fluctuations in their mean batting average are larger than the
systematic decrease in the dispersion.  Thus outliers become rarer and
exceptional achievements, such as a season batting average over .400, or a
consecutive-game hitting streak longer than 56 achieved by Joe DiMaggio also
in 1941, should not recur.

In the next section, we define the rat race model and then we analytically
determine its dynamical features for a two-agent system in Sec.~\ref{2C}.  In
Sec.~\ref{DET}, we investigate many agents in the framework of an almost
deterministic version of the rat race.  Simulation results for the
evolutionary behavior of the model are given in Sec.~\ref{SIM}, and we
conclude in Sec.~\ref{SUM}.

\section{RAT RACE MODEL}
\label{RRM}

\begin{figure}[ht]
\includegraphics*[width=0.35\textwidth]{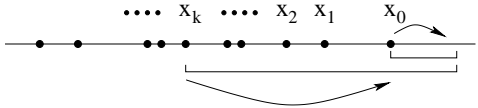}
\caption{Stochastic rat race model.  Each particle has a fitness $x_k$.  The
  leader can advance by an amount that is uniformly distributed in the range
  $x_0-x_1$.  The $k^{\rm th}$ particle can advance by an amount that is
  uniformly distributed in $m(x_0-x_k)$.}
\label{model}
\end{figure}

In our rat race model, each individual $i=0,1,2,\ldots$ possesses a
real-valued fitness $x_i$, with larger $x_i$ representing higher fitness
(Fig.~\ref{model}).  An individual attempts to improve with respect to the
competition by advancing to larger $x$.  Advancement events occur one at a
time and each individual has the same rate of advancing; {\it i.e.}, we
consider {\em serial\/} dynamics in which a randomly-selected competitor
advances.  The leader, located at $x_0$, advances by an amount that is drawn
from a uniform distribution of width $x_0-x_1$.  That is, the leader is aware
only of the next strongest individual and attempts to maintain its lead by
advancing by an amount that is of the order of the separation to this nearest
neighbor.  On the other hand, all other individuals seek to overtake the
leader.  The $i^{\rm th}$ agent, with fitness $x_i$, moves a distance that is
uniformly distributed in the range $m(x_0-x_i)$.  Here $m$ is the fundamental
parameter---the ``catch-up'' factor---that quantifies the severity of the
competition.  When $m<1$, the leader maintains the lead forever, while for
$m>1$ the leader can be overtaken.

In the context of competition, it would be more realistic to eliminate
laggards and replace them by typical individuals.  However, our model mimics
precisely this situation, as a laggard typically moves toward the average
fitness.  A lazy population is characterized by a small value of $m$ for
which the leader maintains the lead on the rest of the pack.  For a
sufficiently large value of $m$, however, the lead changes often and by large
amounts so that the width of the fitness distribution increases after each
advancement event.  Between these two extremes there is an intermediate
regime of stasis where the spread of the pack shrinks to zero and the
population stops advancing.

\section{TWO COMPETITORS}
\label{2C}

We begin by studying the case of two agents with fitnesses $x_0$ and
$x_1<x_0$ and gap $g=x_0-x_1$.  The fitness of the leader increases by an
amount that is uniformly distributed in $[0,g]$ to try to maintain its lead.
Similarly, the laggard advances by a distance that is uniformly distributed
in the range $[0,mg]$.  For $m<1$ the agents always maintain their order,
while for $m>1$ lead changes can occur.  We now determine the evolution of
the gap length for any $m$.

The gap length undergoes a random multiplicative process because each
advancement step leads to a multiplicative change.  Thus we expect that the
distribution of gap lengths for an ensemble of two-agent systems will have a
log-normal form.  Additionally, over a suitable range for the catch-up factor
$m$ we also expect large fluctuations between different realizations of the
process, as is well-known to occur in random multiplicative processes
\cite{rmp}.

\begin{figure}[ht]
  \includegraphics*[width=0.3\textwidth]{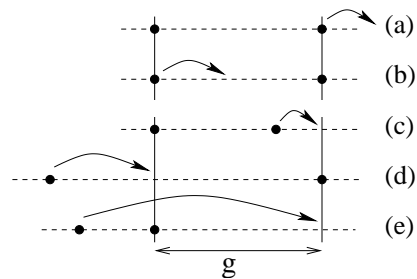}
  \caption{Advancement events that contribute to the change in the gap length
    for a two-particle rat race.  Gaps of length $g$ are lost by the first two
    process, while the latter three lead to a gain of gaps of length $g$.  The
    last process is a lead-changing event.}
\label{gap}
\end{figure}

\subsection{Gap Evolution}

The gap evolution is completely described by $P(g,t)$, the probability for a
gap of length $g$ at time $t$.  For the case $m<1$ (no lead changes), the
evolution of $P$ is described the master equation
\begin{equation}
\label{me<}
\begin{split}
\dot P(g) &= -P(g) +\frac{1}{2}\int_{\frac{g}{2}}^g \frac{P(g')}{g'}\, dg' \\
&~~~~~~~~~~~~~~~~~~+ \frac{1}{2m}\int_{g}^{\frac{g}{1-m}} \frac{P(g')}{g'}\, dg'\,,
\end{split}
\end{equation}
where the overdot denotes time derivative.  The first term on the right
accounts for the loss of gaps of length $g$ because of the hopping of either
particle ((a), (b) in Fig.~\ref{gap}).  The second term accounts for the
creation of a gap of length $g$ due to the leader advancing from a previous
gap of length $g'<g$ (Fig.~\ref{gap}(c)).  The length $g'$ of this previous
gap must be in the range $[g/2,g]$ so that a gap of length $g$ can be created
and the factor $1/g'$ accounts for the hopping distance being uniformly
distributed in $[0,g']$.  The last term accounts for the laggard advancing to
create a gap of length $g$ (Fig.~\ref{gap}(d)).  Here, the previous gap
length $g'$ must be in the range $[g,g/(1-m)]$ and the hopping probability
then equals $1/(mg')$.

Similarly, the master equation for $P(g)$ for $m>1$ is:
\begin{equation}
\label{me>}
\begin{split}
\dot P(g) &= -P(g) +\frac{1}{2}\int_{\frac{g}{2}}^g \frac{P(g')}{g'}\, dg' \\
&~~~~~~+ \frac{1}{2m}\int_{g}^{\infty} \frac{P(g')}{g'}\, dg'
+ \frac{1}{2m}\int_{\frac{g}{m-1}}^\infty \frac{P(g')}{g'}\, dg'\,.
\end{split}
\end{equation}
The third term on the right accounts for events in which the laggard remains
the laggard (Fig.~\ref{gap}(d)), while the last term accounts for overtaking
events (Fig.~\ref{gap}(e)).

For both $m<1$ and $m>1$, it is straightforward to check that these equations
conserve the total probability, $\int_0^\infty \dot P(g)\, dg=0$.  For this
purpose, we need to compute
\begin{equation}
\label{me<-int}
\int_0^\infty \dot P(g)\, dg = \int_0^\infty
\big[\cdots]\,  dg
\end{equation}
where $[\cdots]$ denotes the right-hand side of Eq.~\eqref{me<} or
Eq.~\eqref{me>}.  To perform this type of integral, we merely interchange the
order of the $g$ and $g'$ integrations.  We illustrate this calculation for
the second term on the right-hand side of Eq.~\eqref{me<}.  The interchange
of integration order in this term gives
\begin{eqnarray*}
\int_0^\infty dg\,\, \frac{1}{2}\int_{g/2}^g\frac{P(g')}{g'} dg' 
= \frac{1}{2}\int_0^\infty dg' \int_{g'}^{2g'} \frac{P(g')}{g'}\, dg
\end{eqnarray*}
The integration over $g$ merely gives $g'$ and then the $g'$ integral becomes
simply $\frac{1}{2}\int_0^\infty P(g')\, dg'=\frac{1}{2}$.  The same
manipulation works for all the other terms in the master equation and we thus
verify that $\int P(g)\, dg$ is conserved.

\subsection{Moments of the Gap Length}

The equation of motion for the moments of the gap-length distribution is
\begin{equation}
\label{gav<}
\dot {\mathcal{M}_k}\equiv \left\langle \frac{d g^k}{dt}\right\rangle = 
\int_0^\infty g^k \dot P(g)\, dg =\int_0^\infty 
g^k\, \big[\cdots]\, dg  \,,
\end{equation}
where $[\cdots]$ again denotes the right-hand side of Eq.~\eqref{me<} or
Eq.~\eqref{me>}.  Employing the same interchange of integration order as
illustrated above, the integrals can be evaluated straightforwardly to yield
the following closed equations for the moments:
\begin{eqnarray}
\label{moments}
\dot {\mathcal{M}_k} =  \mathcal{M}_k\!\times\!
\begin{cases}
{\displaystyle  -1 +\frac{2^{k+1}\!-\!1}{2(k+1)} 
+ \frac{1-(1\!-\!m)^{k+1}}{2m(k+1)}} &m<1;\\ \\
{\displaystyle  -1 +\frac{2^{k+1}\!-\!1}{2(k+1)} 
+ \frac{(m\!-\!1)^{k+1}+1}{2m(k+1)}} &m>1.
\end{cases}
\end{eqnarray}

For $m<1$, the first few moments obey:
\begin{eqnarray}
\label{mom<}
\dot {\mathcal{M}}_0 &=& 0 \nonumber \\
\dot {\mathcal{M}_1} &=&   \mathcal{M}_1\left(\frac{1-m}{4}\right)\nonumber \\
\dot {\mathcal{M}_2} &=&   \mathcal{M}_2\left(\frac{3-2m+m^2}{6m}\right)
\end{eqnarray}
{\it etc}.  All  positive integer moments increase in time for $m<1$
because the leader hops further than the laggard, on average, in every single event.
Conversely, for $m>1$ the corresponding moment equations are:
\begin{eqnarray}
\label{mom>}
\dot {\mathcal{M}_0} &=& 0 \nonumber \\
\dot {\mathcal{M}_1} &=&   \mathcal{M}_1\left[\frac{(m-1)(m-2)}{4m}\right]\nonumber \\
\dot {\mathcal{M}_2} &=&   \mathcal{M}_2\left[\frac{(m-1)^3+m+1}{6m}\right]
\end{eqnarray}
{\it etc}.   
Curiously, different moments can have opposite time dependences.  For $m<1$
the laggard trails further and further behind after each step and the average
separation grows, while for $m>2$, overtaking events are so drastic in
character that the average separation between the two agents also grows.
Conversely, for $1<m<2$, the first moment decreases in time.  In spite of the
differing behaviors for the first moment as a function of $m$, higher moments
grow for any $m>1$ (Eq.~\eqref{mom>}).

Why does this dichotomy between moments of different order arise?  The source
is the multiplicative process that underlies the gap dynamics.  This
multiplicativity leads to the very broad {\it log-normal\/} distribution of
gap sizes (to be derived in the next section), for which the time dependence
of moments of different order can be quite different \cite{rmp}.  In a random
multiplicative process, extreme realizations with an exponentially small
probability, make an exponentially large contribution to the moment of a
given order.  For $m<1$ or $m>2$, the interplay between these two extremes
leads to a first moment that grows with time when summing over all
realizations.  In simulations, however, we study only a small fraction of all
realizations and thus can observe only the very different {\it most probable}
behavior.

The most probable gap $g_{\rm mp}=e^{\langle\ln g\rangle}$ (the geometric
average of $g$) may be obtained by computing $\dot X\equiv \langle \dot \ln
g\rangle$, using the same approach that leads from Eq.~\eqref{gav<} to
\eqref{moments}.  We thereby find that $X=At$ with
\begin{eqnarray}
\label{log-moments}
A =  
\begin{cases}
  {\displaystyle \ln 2 -1 +\frac{(m-1)\ln(1-m)}{2m} }&m<1;\\ \\
  {\displaystyle \ln 2 -1 +\frac{(m-1)\ln(m-1)}{2m} }&m>1.
\end{cases}
\end{eqnarray}
Setting $A=0$ gives the transition at which the most probable gap length does
not change.  Again there are two transitions; from the first line of
\eqref{log-moments}, the condition $A=0$ gives a transcendental equation for
$m$ with solution $m_1^* =0.596754\ldots$.  Similarly, from the second line
of \eqref{log-moments}, the condition $A=0$ gives the threshold
$m_2^*=3.38846\ldots$.  The most probable gap length $g_{\rm mp}$ thus
increases with time for $m<m_1^*$ and $m>m_2^*$, while $g_{\rm mp}$ shrinks
to zero in a finite time for $m_1^*<m<m_2^*$.  Comparing Eqs.~\eqref{mom>} \&
\eqref{log-moments}, there exists a range of $m$ for which the average gap
grows while the most probable gap shrinks.  Again, the interplay between
exponentially unlikely events that have exponentially large contributions to
an observable gives seemingly contradictory results that are natural outcomes
of a random multiplicative process \cite{rmp}.

\subsection{The Gap Length Distribution}

We now compute the asymptotic tails of the gap length distribution itself.
Our approach to determine this distribution is to write the moments of the
gap length distribution in Eqs.~\eqref{moments} as a Fourier transform and
then invert this transform.

Thus we write
\begin{eqnarray*}
\int_0^{\infty} P(g,t)\, g^k\, dg = e^{\ln[\mathcal{M}_k(t)]}\,.
\end{eqnarray*}
Now define $X= \ln g$ and make an analytic continuation from $k$ to $ik$ to
give
\begin{eqnarray*}
\int_{-\infty}^{\infty} P(X,t) \,e^{i k X}\, dX = e^{\ln[\mathcal{M}_{ik}(t)]}\,.
\end{eqnarray*}
The left-hand side is just the Fourier transform of $P(X,t)$.  Inverting this
Fourier transform, we obtain
\begin{eqnarray*}
P(X,t) = \frac{1}{2\pi}\int_{-\infty}^{\infty} dk \,\,e^{-ik X} \,e^{\ln[\mathcal{M}_{ik}(t)]}\,.
\end{eqnarray*}
To derive the asymptotic distribution for large $X$, we need the small-$k$
behavior of $\ln \mathcal{M}_{ik}$.  Using \eqref{moments}, we expand $\ln
\mathcal{M}_{ik}$ for small $k$ and then invert the Fourier transform to
obtain a Gaussian distribution for $X$, {\it i.e.}, a log-normal distribution
for $g$.  The final result is
\begin{eqnarray}
P(X,t) \sim \frac{1}{\sqrt{2\pi B t}}\,e^{-(X-A t)^2/{2B t}}\,,
\end{eqnarray}
where $A$ is given by Eq.~\eqref{log-moments} and
\begin{eqnarray}
B=
\begin{cases}
 C - \frac{m-1}{2m}\left[2\ln(1-m) - \ln^2(1-m)\right]& m < 1 \\ \\
 C - \frac{m-1}{2m}\left[2\ln(m-1) - \ln^2(m-1)\right]&  m > 1,
\end{cases}
\end{eqnarray}
with $C=2-2\ln 2 +(\ln 2)^2$.  We again emphasize that while the distribution
of $X=\ln g$ extends over range that grows as $\sqrt{t}$, the distribution of
$g$ itself is extremely broad so that it cannot be characterized by any
individual moment.

\section{DETERMINISTIC MODEL}
\label{DET}

It is not clear how to adapt the theory given above in an
analytically-tractable way to treat more than 2 particles.  We therefore
introduce an nearly-deterministic version of the model that mimics the
advancement steps in the stochastic rat race model by defining the length of
each jump to be exactly one half of the total possible range.  We again
consider serial dynamics in which one of the competitors, chosen at random,
advances.  The order in which competitors are selected is the only source of
stochasticity in this version of the model.

\subsection{Two Particles}
\label{N1}

There are two possibilities for particle movement, depending on the value of
the catch-up factor $m$:
\begin{itemize}
\item For $m<2$, if the leader moves, the gap $g\to(3/2)g$, while if the
  laggard moves, the gap $g\to \beta g$, where $\beta=1-(m/2)$.
\item For $m>2$ the laggard overtakes the leader.  If the leader moves, again
  $g\to(3/2)g$, while if the laggard moves $g\to \beta g$, where
  $\beta=(m/2)-1$.
\end{itemize}
Since either particle is selected with probability 1/2 at each step, after
$t$ steps the gap could assume any of the values
$(3/2)^{\tau}\beta^{t-\tau}$, $\tau=0,1,\dots,t$ (assuming an initial gap
length $g=1$).  After $t$ steps, the probability of a gap of length
$(3/2)^{\tau}\beta^{t-\tau}$ is
\[
p_{\tau}=\frac{1}{2^t}\Big({t\atop\tau}\Big)\,.
\]
\begin{figure}[ht]
\includegraphics*[width=0.45\textwidth]{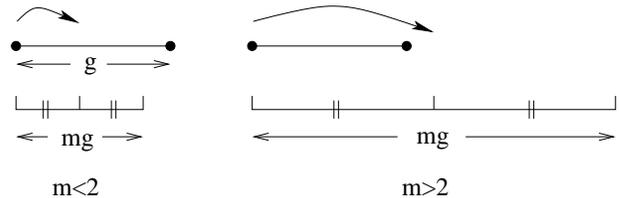}
\caption{Deterministic rat race.  The laggard advances by $mg/2$.  For $m<2$,
  the order never changes, while for $m>2$, the move of the laggard always
  leads to passing.}
\label{model-det}
\end{figure}

It follows that the $k^{\rm th}$ moment of the gap length is
\begin{equation}
\label{Mnu}
\mathcal{M}_{k}\equiv\av{g^{k}}=\Big[\frac{1}{2}\Big(\frac{3}{2}\Big)^{k}+\frac{1}{2}\beta^{k}\Big]^t\,.
\end{equation}
Thus, the large-time behavior of the $k^{\rm th}$ moment depends on the
factor $[(3/2)^k+\beta^k]/2$.  If this factor is greater than $1$, {\it
  i.e.}, $\beta$ exceeds $\beta_c(k) =\exp[\ln(2-(3/2)^k)/k]$, the $k^{\rm
  th}$ moment diverges as $t\to \infty$.  On the other hand, for
$\beta<\beta_c(k)$, the $k^{\rm th}$ moment vanishes as $t\to \infty$.  For
$\beta>\beta_c(0)=2/3$, {\it all\/} positive moments of the gap size, as well
as the most probable gap size, diverge.  The value $\beta_c(0)=2/3$ thus
marks the transition from convergent to explosive behavior in the gap size.
Notice that this transition value for $\beta$ corresponds to the threshold
values $m_1^*=2/3$ and $m_2^*=10/3$, which agree well with the corresponding
thresholds from the stochastic rat race.

\subsection{Many Particles}

We study an $N+1$-particle system, with $N>1$, with particles located at
$x_i$, with $i=0,1,2,\ldots N$.  The gap between particle $n$ and the leader
is defined as $g_n=x_0-x_n$.  We limit ourselves to the case of catch-up
factor $m>2$, so that any non-leader that jumps always overtakes the leader.
This fact allows us to keep track of the ordering of the particles and an
exact analysis is then possible.  For generality, we assume the leader jumps
a distance $\alpha g_1$ ahead; $\alpha=1/2$ corresponds to the case analyzed
previously, while for $\alpha=0$ the leader is completely lazy and never
jumps.

If particle $n$ is selected, this results in a re-distribution of the vector
$(g_1,g_2,\dots,g_N)$ of the gap lengths:
\[
\g'=\A_n\g\,,
\]
where 
\[ 
\A_0=  
\left (  
\begin{array}{ccccc}  
1+\alpha & 0 & 0 &\cdots & 0 \\  
1-\alpha & 1 & 0 &\cdots & 0 \\
1-\alpha & 0 & 1 &\cdots & 0 \\
\vdots&\vdots&\vdots&\ddots&\vdots\\
1-\alpha&0&0&\cdots&1
\end{array}\right),\>
\A_1=  
\left (  
\begin{array}{ccccc}  
\beta & 0 & 0 &\cdots & 0 \\  
\beta & 1 & 0 &\cdots & 0 \\
\beta & 0 & 1 &\cdots & 0 \\
\vdots&\vdots&\vdots&\ddots&\vdots\\
\beta & 0&0&\cdots&1
\end{array}    \right), \]
\[
\A_2=  
\left (  
\begin{array}{ccccc}  
0& \beta  & 0 &\cdots & 0 \\  
1&\beta  & 0 &\cdots & 0 \\
0&\beta  &1 &\cdots & 0 \\
\vdots&\vdots&\vdots&\ddots&\vdots\\
0&\beta &0&\cdots&1
\end{array}    \right),  \dots\>
\A_N=  
\left (  
\begin{array}{ccccc}  
0& 0  & 0 &\cdots & \beta \\  
1&0  & 0 &\cdots & \beta \\
0&1  &0&\cdots & \beta \\
\vdots&\vdots&\ddots& &\vdots\\
0&0 &\cdots&1&\beta
\end{array}    \right).
\] 
The vector \g, after $t$ steps, is the result of the product of $t$ matrices,
drawn at random from among the $\A_n$.  Unfortunately, none of the matrices
commute, so we cannot deduce the probability distribution of gap sizes as in
the scalar case of two particles ($N=1$).  On the other hand, according to
the Oseledec theorem \cite{O} (also known as the multiplicative ergodic
theorem), the growth of \g\ is determined by the product
$\tlam_0\tlam_1\cdots\tlam_N$, where $\tlam_n$ is the largest eigenvalue of
the matrix $\A_n$ ($n=0,1,2,\dots N$).  More precisely, the product
$\mathcal{P}\equiv \prod_{i=0}^N \tilde \lambda_i$ determines the largest
Lyapunov exponent of the growth of \g\ with $N$.  Thus explosive growth
results if the product $\mathcal{P}>1$, and stasis results otherwise.  We now
derive the critical value for $\beta$ at the transition point, where
$\tlam_0\tlam_1\cdots\tlam_N=1$.

The largest eigenvalue of $\A_0$ is clearly $\tlam_0=1+\alpha$.  To determine
the largest eigenvalue of $\A_n$ for $n>0$, we write the characteristic
equation det$(\A_n-\lambda{{\bf I}})=0$, obtained by expanding the
determinant about the column of $\beta$'s:
\[
[\beta(1+\lambda+\lambda^2+\cdots+\lambda^{n-1})-\lambda^n](1-\lambda)^{N-n}=0\,.
\]
From the second factor on the left-hand side we conclude that $\A_n$ has
$N-n$ eigenvalues $\lambda_n=1$.  We argue, self-consistently, that for
$n=1,2,\dots,N-1$ these are also the largest eigenvalues, that is,
$\tlam_n=1$.  Indeed, if that is the case, then the criticality condition
dictates $\tlam_N=1/(1+\alpha)$.  Substituting this value into the
characteristic equation for $\A_N$ we find
\begin{equation}
\label{bc_a>0}
\beta^*=\frac{\alpha}{(1+\alpha)^{N+1}-(1+\alpha)}\,.
\end{equation}
It follows that $\beta^*\leq1/N$ (with the equality being realized in the
limit $\alpha\to0$).  We can now show that our initial assumption that the
remaining eigenvalues of $\A_n$ ($n=1,2,\dots,N-1$) are not larger than 1 is
indeed valid.  From the factor in the square brackets of the characteristic
equation, we see that these eigenvalues satisfy
\[
\beta=\frac{\lambda^n}{1+\lambda+\lambda^2+\cdots+\lambda^{n-1}}\,.
\]
The expression on the right-hand side is a monotonically increasing function
of $\lambda$, hence if $\lambda>1$ then $\beta>1/n>1/N$, in contradiction
with~(\ref{bc_a>0}).

For $\alpha=0$, namely, the case of a lazy leader that never advances as long
as it leads, the critical value of the catch-up parameter $\beta$ is no
longer exponential in $N$, but rather
\begin{equation}
\label{bc_a=0}
\beta^*=\frac{1}{N}\,,
\end{equation}
as can be seen by taking the limit of $\alpha\to0$ in  Eq.~(\ref{bc_a>0}).  

\section{SIMULATION RESULTS}
\label{SIM}

For any number of agents, the time dependence of the fitness of each agent
exhibits rich behavior.  For 2 agents, simulations clearly show a transition
between a regime where the leader runs away from the laggard and stasis as
$m$ passes through a critical value close to $m_1^* =0.596754\ldots$.  This
stasis continues until a second transition at $m\approx m_2^*=3.38846\ldots$.
For $m> m_2^*$, there is explosive growth, with many lead changes between the
two agents.  It bears emphasizing that the observed transitions occur close
to the values associated with the most probable gap size, even though the
true transitions occur at $m_1=1$ and $m_2=2$, corresponding to the average
gap size.  Since simulations reflect the most probable behavior, they can
provide qualitative information about the nature of stasis and explosive
growth, as well as the transition between these two regimes, but little else.

\begin{figure}[ht]
  \includegraphics*[width=0.36\textwidth]{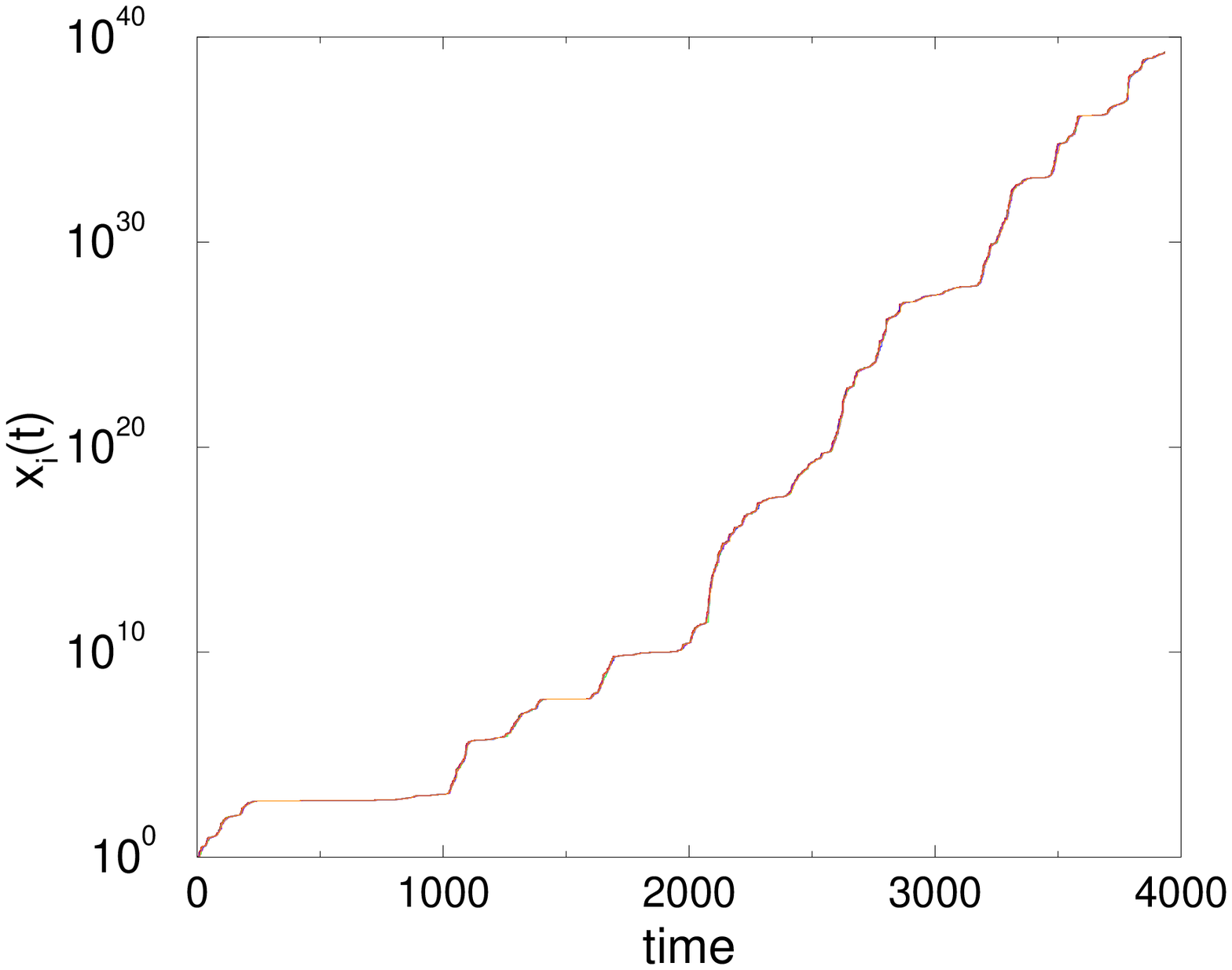}
  \includegraphics*[width=0.33\textwidth]{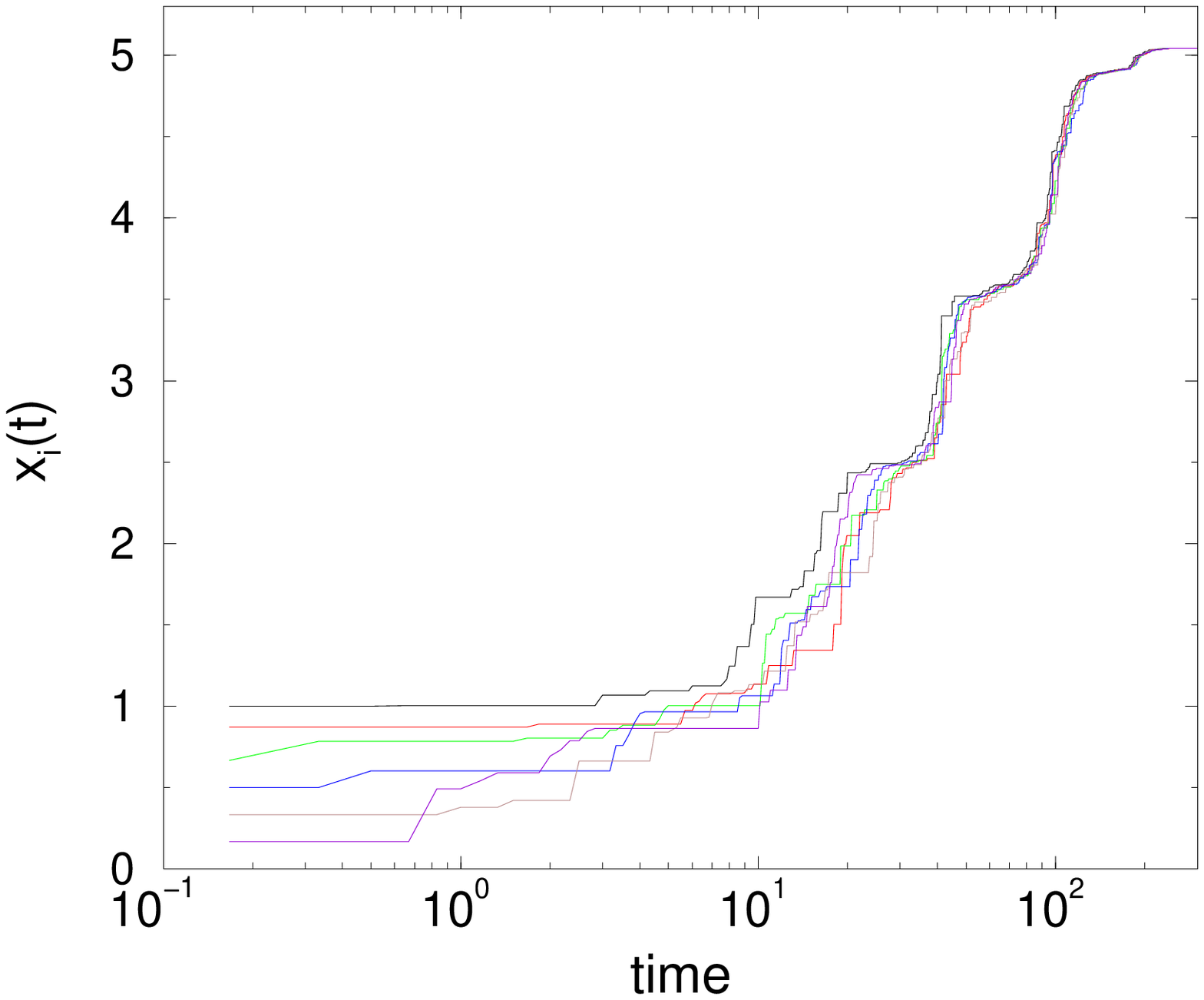}
  \caption{Fitnesses of each agent for a 6-particle system for $m=0.4$ on a
    linear-log scale (top), and $m=0.44$ on a log-linear scale (bottom).  }
\label{N6a}
\end{figure}

\begin{figure}[ht]
  \includegraphics*[width=0.36\textwidth]{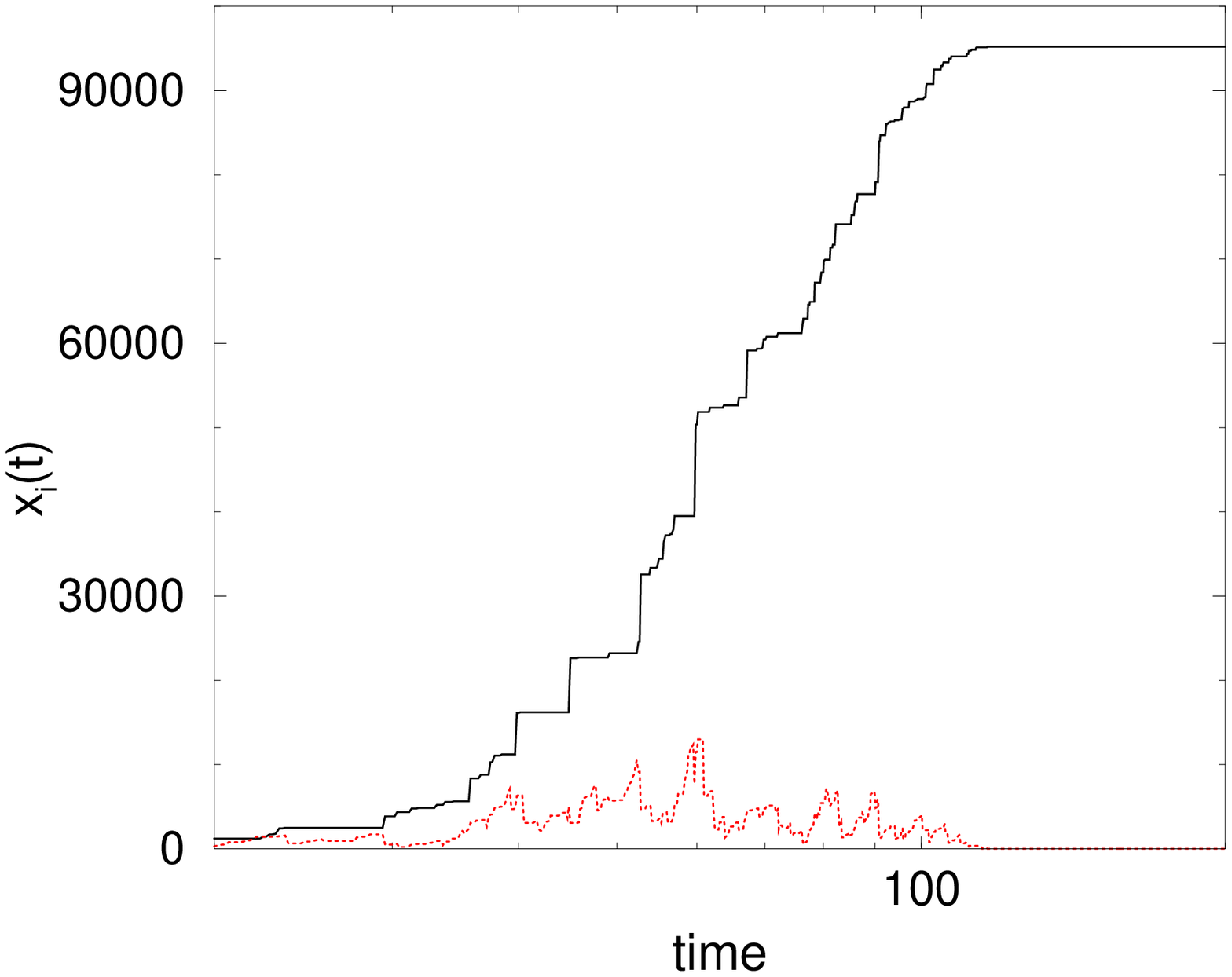}
  \includegraphics*[width=0.375\textwidth]{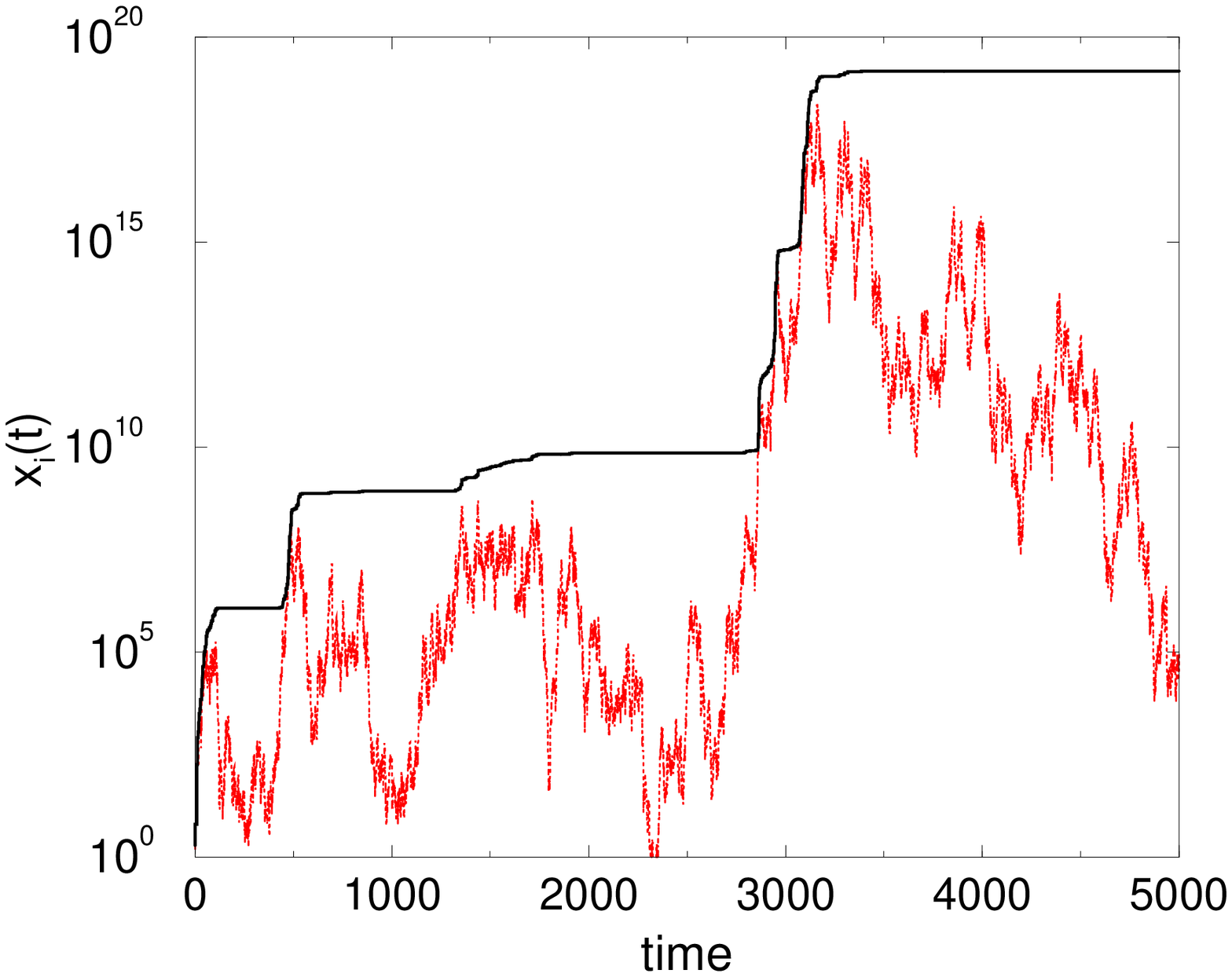}
  \caption{Fitnesses of each agent for a 6-particle system for $m=1.61$ on a
    log-linear scale (top) and $m=1.64$ on a linear-log scale (bottom).
    Shown dotted is the distance between the strongest and weakest agent. }
\label{N6b}
\end{figure}

Figs.~\ref{N6a} and ~\ref{N6b} show typical results for 6 agents.  Again, the
existence of two transitions is clearly visible.  For $m<m_1^*\approx 0.4$,
the initial leader always maintains its lead, but the laggards are able to
remain relatively close behind by virtue of the multiplicative nature of the
jumps.  Strikingly, large jumps occur with some frequency so that the
population still advances rapidly.  However, for slightly larger $m$, the
distance between the strongest and weakest eventually disappears and the
evolution quickly grinds to a halt (Fig.~\ref{N6a} bottom).  Here lead
changes are rare and no longer occur after  a short time.  This nearly
static behavior continues until $m\approx 1.6$ (Fig.~\ref{N6b} top).

For a slightly larger $m$, this system exhibits periods of near stasis
followed by periods of explosive growth (Fig.~\ref{N6b} bottom).  Here lead
changes occur at roughly a constant rate and the total number of lead changes
grows linearly with time.  During periods of rapid advancement, the gap
between the strongest and weakest agent is nearly comparable to the fitness
(position) of any agent.  Conversely, during periods of near stasis, the gap
between the strongest and weakest agent is orders of magnitude smaller than
the typical fitness.

Qualitatively similar behavior occurs for more particles, except that the
critical values of $m$ that separates leader runaway from stasis and stasis
from explosive growth seem to approach 0 and 1, respectively, as the number
of particles increases.

\section{SUMMARY}
\label{SUM}

We introduced an idealized social competition model where individuals try to
better their fitness (a real-valued variable) by advancing relative to the
rest of the population: the leader advances to remain ahead of its closest
pursuer, while others in the pack advance by a random amount, in proportion
to their distance from the leader.  When this proportionality constant is too
small or too large, the fitness of all the agents grows explosively during
short sporadic bursts.  These explosive regimes are the analog of the rapid
growth of new species after massive die-offs in punctuated evolution.
Between these two extremes there is a window of stasis, where the spread of
the pack shrinks indefinitely and evolution comes to a stop.  Here outliers
becomes progressively less likely and extreme achievements disappear; this
situation parallels the disappearance of the .400 hitter in baseball
mentioned in the introduction.

Basic features of the model already arise in the simple limits of just two
agents, and in a deterministic model where agents advance by a fixed multiple
of their gap to the leader.  These simplified models allow for an exact
analysis, yielding specific expressions for the distribution of gaps in the
two-agent model, and for the $N$-dependence of the threshold parameters that
demarcate between the regimes of stasis and explosive growth in the
deterministic model.  Our simulation results suggest that similar behavior
occurs for a general many-agent system.  A full analytical solution of the
general many-agent problem seems intractable, however, even in the simplified
deterministic version.  Thus some basic questions remain unanswered, such as,
for example, what is the distribution of agents in the pack in the various
regimes of explosive growth, stasis, and at the critical transition points.

\noindent{\bf Acknowledgments.} We thank the Isaac Newton Institute for
Mathematical Sciences (Cambridge, England), where this research was started,
for its hospitality.  Two of us gratefully acknowledge financial support from
NSF grant PHY0555312 (DbA) and DMR0535503 (SR).

\end{document}